\documentclass[]{spie}  
\usepackage[]{graphicx}
\usepackage{html}
\usepackage{hyperref}

\title{AMBER closure and differential phases: accuracy and calibration with a Beam Commutation\thanks{Based on ESO paranal observations of calibrators made with AMBER
    during various commissioning and GTO runs}}

\author{Florentin Millour\supit{a},
  Romain G. Petrov\supit{b},
  Martin Vannier\supit{b}
  and  Stefan Kraus\supit{a}
  \skiplinehalf
  \supit{a}Max-Planck Institut for Radioastronomy, Auf dem h\"ugel, 69,
  53121, Bonn, Germany; \\
  \supit{b}Fizeau Laboratory, Nice university, Parc Valrose, Nice,
  France.
}

\authorinfo{Further author information: (Send correspondance to
  F. Millour)\\F. Millour: E-mail: fmillour@mpifr-bonn.mpg.de,
  Telephone: 49 228 525 188}

\begin{document}
\maketitle


\begin{abstract}
  The first astrophysical results of the VLTI focal instrument AMBER
  have shown the importance of the differential and closure phase
  measures, which are supposed to be much less sensitive to
  atmospheric and instrumental biases than the absolute
  visibility. However there are artifacts limiting the accuracy of
  these measures which can be substantially overcome by a specific
  calibration technique called Beam Commutation. This paper reports
  the observed accuracies on AMBER/VLTI phases in different modes,
  discusses some of the instrumental biases and shows the accuracy
  gain provided by Beam Commutation on the Differential Phase as well
  as on the Closure Phase.
\end{abstract}


\keywords{Instrumentation: interferometers
  - Techniques: image processing
  - Methods: data analysis
}

\section{Introduction}
\label{sec:intro}  

In Optical long baseline interferometry, the phase of the object
complex visibility is lost because of the random and fast phase variations introduced by the atmosphere.
Three main techniques exist to
overcome this strong limitation and recover a part of the phase
information: phase referencing\cite{1998SPIE.3350..807Q}, where
an off axis source in the isoplanetic angle is used as a phase
reference, closure phase\cite{1986Natur.320..595B,
  2005AJ....130..246K}, which is a combination of the phases from
three telescopes and is by nature free of any disturbance introduced
in any of the beams, and then differential
phase\cite{1999ASPC..194...89A, 2006MNRAS.367..825V}, which is the
same as phase referencing but with the source itself in some particular
wavelength as a phase reference.

The phase referencing technique is affected by atmospheric effects, such
as chromatic optical path difference (OPD), which is not exactly the
same for the reference and the science target. Instrumental effects
also affect the phase. Phase referencing technique uses a high
precision metrology and a technique of commutation between the science
star and the off axis star to overcome these effects.
The OPD effects completely disappear in the phase closure, which
explain why this phase-measurement technique has been used up to now
with great success. However, for high dynamic and high precision
measurements, closure phase remains sensitive to all non-OPD
instrumental effects.
The differential phase is affected by the same effects as the phase
referencing technique. The instrumental effects can be reduced
by a careful instrumental design, making the conditions of measure of
all spectral channels (for example) as identical as possible, but they
are almost impossible to eliminate completely. Therefore, a
commutation technique would allow one, just like for phase
referencing, to overcome in great part the instrumental effects.

To calibrate the differential and closure phases of AMBER, we
installed a specific tool called the Beam Commutation Device (BCD). It
has been inspired by our experience in differential speckle
interferometry (Cuevas and Petrov, 1994) \cite{1994SPIE.2200..501C}
and consists in commuting two of the input beams without
otherwise modifying the images pupils positions and OPD values seen by
the instrument. The main advantage of such a commutation is that it
can be performed very rapidly, to the contrary of usual calibration
techniques. 

This paper presents a first set of tests of this calibration method on
the AMBER instrument and compares it results with those of a standard
calibration using a reference star observed at a different time than
the science source.

\section{AMBER phases and Beam Commutation}

We first remind the definitions of differential and closure phase,
then discuss their possible biases, the standard calibration
procedure, using a reference star and the BCD calibration procedure.

\subsection{Differential phase and Closure phase measures and perturbations}

\subsubsection{The differential phase}

In optical, as well as in radio astronomy, source phase information
refers to the phase of its complex visibility $\mu$. In a single-mode
interferogram, the phase is related to the position of the fringes,
and in the absence of nanometer accuracy metrology, the measured phase
$\phi_{12}^{m}(\lambda)$ is affected by an unknown instrumental term
linked to the VLTI and AMBER differential piston $\delta_{12}(t)$ and to the
instantaneous atmospheric piston $p_{12}(t)$ between beams $1$ and $2$, as
a function of the wave number $\sigma=1/\lambda$:

\begin{equation}
  \phi^{m}_{12}(t,\sigma) = \phi^{*}_{12} (\sigma) + 2 \pi \left[
    \delta_{12}(t) + p_{12}(t) \right] \sigma
\end{equation}

At this point, we can introduce a development of the source phase
$\phi_{12}^{*}(\sigma)$ that will be used afterwards:

\begin{equation}
  \phi_{12}^{*}(\sigma)=a^{*}_{0}+a^{*}_{1}\sigma +
  \delta\varphi_{12}^{*}(\sigma)
\label{eq:dvptPhase}
\end{equation}

where $\delta\varphi_{12}^{*}(\sigma)$ stands for the object phase
that cannot be explained by a phase slope as a function of wave
number. It contains higher-order terms of the phase polynomial
development that are the source information which can be extracted
from differential phase measurements.

To compute the differential phase in practice, we use the so-called
complex coherent flux, which is usually the rawest measure an
interferometer outputs. It contains the instantaneous complex
visibility $\mu$, the instantaneous flux $f$, and an additional term
$T$ that we call ``transfer function'' which contains all the
atmospheric and instrumental effects on the complex visibility:

\begin{equation}
  C^m_{12}(t, \sigma) = F(t, \sigma) T(t, \sigma) \mu(\sigma)
\end{equation}

The phase of the transfer function term $T$ contains the terms
$\delta_{12}(t)$ and $p_{12}(t)$ described before, and the phase of $\mu$
contains the object phase $\phi_{12}^{*}(\sigma)$. The differential
phase is obtained in practice by fitting the measured phase
$\phi^{m}_{12}(t, \sigma)$ with a linear function $\varphi^{\rm fit}_{12}$ of
$\sigma$, therefore enclosing the terms $\delta_{12}$, $p_{12}$,
$a^{*}_{0}$ and $a^{*}_{1}$:

\begin{equation}
  \varphi^{\rm fit}_{12}(t, \sigma) = 2\pi(\delta_{12}(t) + p_{12}(t)
  + a^{*}_{1})\sigma + a^{*}_{0}
\label{eq:fittingPhase}
\end{equation}

We correct then $C^m_{12}(t, \sigma)$ from this effect on a
frame-by-frame basis, obtaining $C^{nop}_{12}(\sigma_{i})$, and we
integrate it in a reference channel, obtaining $C^{\rm ref}_{12}(t,
\sigma_{i})$:

\begin{eqnarray}
  C^{\rm nop}_{12}(t, \sigma_{i}) & = & C^m_{12}(t, \sigma_{i}) e^{-i
    \varphi^{\rm fit}_{12}(\sigma_{i})}\\
  C^{\rm ref}_{12}(t, \sigma_{i}) & = & \sum_{j \neq i}^{n_{\sigma}}
  C^m_{12}(t, \sigma_{j}) e^{-i \varphi^{\rm fit}_{12}(t, \sigma_{j})}
\end{eqnarray}

The differential phase is then computed by:

\begin{equation}
  \delta\phi^m_{12}(\sigma) = arg\left[ < C^{\rm nop}_{12}(t, \sigma)
    \times \left( C^{\rm ref}_{12}(t, \sigma) \right)^{*} >_t \right]
\end{equation}

Therefore, in the way it is defined, this differential phase is an
estimator of the beforehand mentioned (eq.~\ref{eq:dvptPhase}) high
order term of the object phase:

\begin{equation}
  \delta\phi^m_{12}(\sigma) \equiv \delta\varphi_{12}^{*}(\sigma)
\end{equation}






\subsubsection{The effects affecting the differential phase}
\label{sbsect:effects}

The measured differential phase is affected by the source phase
$\delta\varphi^{*}_{12}(\sigma)$, the atmosphere chromatic phase
$\delta\phi^{a}_{12}(\sigma)$, and the instrument chromatic phase
$\delta\phi^{i}_{12}(\sigma)$. It can be expressed as:

\begin{equation}
  \delta\phi^m_{12}(\sigma) = \delta\varphi^{*}_{12}(\sigma) +
  \delta\phi^{a}_{12}(\sigma) + \delta\phi^{i}_{12}(\sigma)
  \label{eq:diffPhase}
\end{equation}

The term $\delta\phi^{a}_{12}(\sigma)$ results only from a chromatic
OPD term as the rapidly variable achromatic term is removed during the
fitting process (eq.~\ref{eq:fittingPhase}). This effect is supposed
afterwards to have slow variations with time. The term
$\delta\phi^{i}_{12}(\sigma)$ contains only the chromatic instrumental
effects, for the same reasons as before. The main effects affecting
the AMBER differential phase are described briefly in the following
list:
\begin{itemize}
\item The atmospheric chromatic effect comes mainly from the different
  thickness of dry and wet air in each beam. Since the refraction
  index of the gases is wavelength dependant, this result in a
  chromatic OPD. A first order correction is made from computed gas
  indexes and known delay line positions, but short term variations of
  temperature, pressure and humidity leave phase terms which vary of
  several hundreds of radians. This effect is linked to dry air and
  water vapour content and is explained in details in Mathar et 
  al.\cite{2004ApOpt..43..928M}. At $2\mu$m, in the AMBER data, this
  effect is seen as a global curvature of the differential phase
  relative to wavelength that can have amplitudes of up to 1\,rad.
\item The dichroics contained in AMBER, due to reflexions and
  refractions, produce a chromatic phase bias seen mainly at the
  beginning of the K band (1.95$\mu$m) and which seems to be stable
  with time.
\item The detector gain table change with time (or the change of the
  pixels used to read the interferogram) produce in first
  approximation and additive effect on the differential phase, which
  can be different from any spectral channel (i.e. lines of
  pixels). This results in a noise of the differential phase as a
  function of wavelength, that might be overcome if the
  time-variations of this $\lambda$-noise are slow enough
\item There are variable modulations in wavelength in the measured
  spectra, visibility, differential phase and closure phase. There are
  some time nicknamed ``socks''. We first suspected the optical fibers
  to be imperfect spatial filters. It has been recently shown that any
  plane parallel plate introduced in the beam actually produces a
  Perot-Fabry effect, with a mixing of waves directly transmitted and
  of one or several waves who have crossed the plate $2n$ more
  times. The most recent measures \cite{Chelli2008}, show that the
  strongest effect, by far, results from the AMBER polarizers put in
  each beam. The effect change with temperature and mechanical
  excitation and result in the variation described below. These
  polarizers will be replaced soon, but this paper shows how the BCD
  cope with these modulations in the differential phases so far.
\end{itemize}

\subsubsection{The closure phase}

The closure phase between baselines \overrightarrow{B_{12}},
\overrightarrow{B_{23}} and \overrightarrow{B_{31}} is the phase of
the average "bi-spectral product" of the coherent fluxes:

\begin{equation}
  \psi_{123}=arg\left[<C^m_{12} \, C^m_{23} \, (C^m_{31})^{*}>_t\right]
\end{equation}

and it is a very classical property of long-baseline interferometry
that this closure phase is independent of any OPD terms affecting
individually each beam. This includes the achromatic OPD, as well
as the chromatic one:

\begin{equation}
  \psi_{123} = \phi^{*}_{12} + \phi^{*}_{23} - \phi^{*}_{13} +
  \phi^d_{12} + \phi^d_{23} - \phi^d_{13}
\end{equation}

where $\phi^d_{ij}$ are error terms linked to the baseline $ij$, due
for example to a change in the detector gain table between the flat
field evaluation and the observation.

\subsubsection{The effects affecting the closure phase}

By construction the closure phase if free from any path difference
introduced in any individual beam. However it is sensitive to
detection effect related to each baseline: 

\begin{itemize}
\item If the pixels (or their gain) used to analyze a spectral channel
  have changed since the instrument calibration, all phases will be
  modified in a way which depends from the baseline interferogram
  calibration and not to the OPD and does not have any reason to
  produce null closure phases. It will appear, as in the case of
  differential phase, as a $\lambda$-noise, that hopefully does not
  vary rapidly with time.
\item The situation with the wavelength beatings and the spatial
  filtering defects are unclear at this point. They introduce phase
  differences in each beam (which should be canceled out in the
  closure phase) but also in each interferogram fit and the latest
  should produce a non zero closure phase as seen below in the
  measures.
\end{itemize}

\subsection{Calibrating the phases with a reference star}
\label{sect:calibRefStar}

\subsubsection{Calibrating the differential phases}

Calibration of the differential phase is the removal of all the
instrumental and atmospheric effects from the phase signal described
in Sect.~\ref{sbsect:effects}. The calibrated differential phase is
thus:

\begin{equation}
  \delta\phi^c_{12}(\sigma) = \delta\phi^{m}_{12}(\sigma) -
  \widehat{\delta\phi}^{a}_{12}(\sigma) +
  \widehat{\delta\phi}^{i}_{12}(\sigma)
  \label{eq:calibDiffPhase}
\end{equation}

where $\widehat{\delta\phi}^{a}_{12}(\sigma)$ and
$\widehat{\delta\phi}^{i}_{12}(\sigma)$ are estimations of the
atmosphere and instrumental differential phases. A way of estimating
these terms is to observe a centro-symmetric star whose phase
$\widehat{\delta\phi}^{*c}_{12}(\sigma)$ is known to be zero at all
wavelengths. The calibrated differential phase is then:

\begin{equation}
  \delta\phi^c_{12}(\sigma) = \delta\phi^{m}_{12}(\sigma) -
  \delta\phi^{*c}_{12}(\sigma)
  \label{eq:calibDiffPhase}
\end{equation}

However, one can see that the calibration must be rapid enough to
consider that the calibration star differential phase
$\delta\phi^{*c}_{12}(\sigma)$ is equal to the differential phase
$\widehat{\delta\phi}^{a}_{12}(\sigma) +
\widehat{\delta\phi}^{i}_{12}(\sigma)$ we want to remove from the
science star. Also, a very similar air mass should be used between the
science star and the calibration star in order to avoid large changes
in the term $\delta\phi^{a}_{12}(\sigma)$. In practice, at VLTI, the
calibration cycle is at least 30\,mn, where the atmospheric water
vapour content has enough time to change by large amounts. Also, in
30\,mn, the instrument has enough time to move as the focal laboratory
temperature slightly changes with time.

\subsubsection{Calibrating the closure phase}

For the closure phase terms linked to a drift of the instrument
calibration, the same technique and restriction applies. They can be
cancelled by the use of a reference star if they are stable other the
20 to 30 mn time necessary to switch from science to reference.

\subsection{Calibrating the phases with beam commutation}

The beam commutation, as briefly explained in introduction, is meant
to remove a part of the phases artifacts presented in
Sect.~\ref{sbsect:effects}. It can be fully seen as an alternative
calibration technique to the standard science-reference stars
presented in Sect.~\ref{sect:calibRefStar}. Indeed, the instrumental
phase effects can be considered stable over a given time span. In
AMBER, for example, the described dichroic phase effect and the
detector gain table are stable over several minutes.

\begin{figure}[htbp]
  \begin{center}
    \includegraphics[width=14cm]{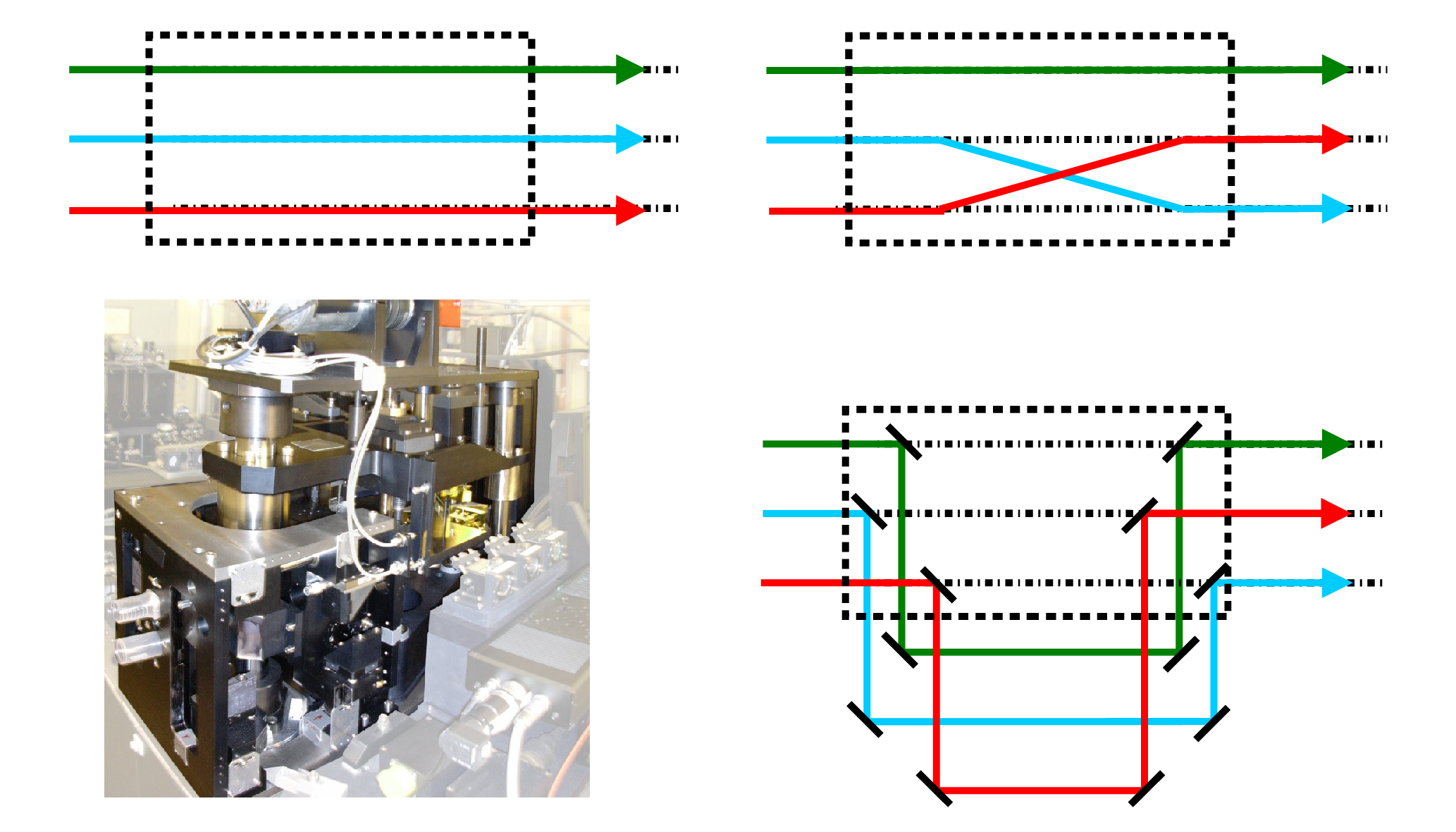}
  \end{center}
  \caption
  { \label{fig:BCDprinciple}
    The beam commutation explained in few sketches:
    \emph{Top-left:} the standard scheme where the beams go through
    unaffected in the instrument.
    \emph{Top-right:} The Beam Commuting Device is inserted in the
    beam and switches two out of the three beams.
    \emph{Bottom-left:} A photo from the authors of the AMBER BCD, as
    it has been build, with 6 additional specular reflexions per beam.
    \emph{Bottom-right:} A simplified sketch of how the AMBER BCD
    commutes the beams. One needs 6 reflexions on flat mirrors to keep
    both the pupil and the image at the same position as well as the
    same OPD.
  }
\end{figure}

\subsubsection{Calibrating the differential phases by beam commutation}

Let us consider that at some point of the interferometer, one can swap
beams 1 and 2 in a specific device called Beam Commuting Device
(hereafter called BCD, see Fig.~\ref{fig:BCDprinciple}). The phase
$\delta\varphi_{12}^{*}(\sigma) + \delta\phi^{a}_{12}(\sigma)$, before
the BCD, is now recorded as being 21 instead of 12, thus changing its
sign compared to the unswapped situation. The phase introduced after
the BCD point $\delta\phi^{i}_{12}(\sigma)$, in the instrument, is
left unchanged:

\begin{eqnarray}
  \mbox{No BCD:} & \delta\phi^{m}_{12}(\sigma)  = & 
  \ \delta\varphi^{*}_{12}(\sigma) + \delta\phi^{a}_{12}(\sigma) +
  \delta\phi^{i}_{12}(\sigma) \nonumber \\
  \mbox{With BCD:} & \delta\phi^{m{\rm,BCD}}_{12}(\sigma)  = & -
  \delta\varphi^{*}_{12}(\sigma) - \delta\phi^{a}_{12}(\sigma) +
  \delta\phi^{i}_{12}(\sigma) + \delta\phi^{\rm BCD}_{12}(\sigma)
  \label{eq:calibDiffPhase}
\end{eqnarray}

The idea is then to record data as usual, to perform the beam
commutation and then to record commuted data. When subtracting the two
measurements, and if the instrument is stable enough (i.e. both the
instrumental term and the atmospheric one are stabilized after the
BCD), the terms $\delta\phi^{a,{\rm A}}_{12}(\sigma) +
\delta\phi^{i,{\rm A}}_{12}(\sigma)$ cancel out:

\begin{equation}
  \delta\phi^c_{12}(\sigma)  =  \frac{\delta\phi^m_{12}(\sigma) -
    \delta\phi^{m,BCD}_{12}(\sigma)}{2} =  \delta\phi^{*}_{12}(\sigma) +
  \delta\phi^{a{\rm B}}_{12}(\sigma) + \delta\phi^{i{\rm
      B}}_{12}(\sigma)
  \label{eq:calibDiffPhase}
\end{equation}

If a beam commuting device is built, it must not affect the beams
with more effect on the phases than the photon noise over one
calibration cycle, i.e. the term $\delta\phi^{\rm BCD}_{12}(\sigma)$
is smaller than the noise on the phase due to fundamental noises
only. One of the main interests of the beam commutation is that it
allows shortening the calibration cycle with regard to the use of a
natural reference star. For example, on AMBER, we go from about 30
minutes to about 2-3 minutes.

One can also see that, using this technique, the term before the BCD
$\delta\phi^{a}_{12}(\sigma)$ cannot be removed. Therefore, one still
has to cope with the atmospheric term, which is anyway present also
when calibrating with a natural reference star.

\subsubsection{Calibration of Closure phase by beam commutation}
\label{sbsect:CPCalib}

\begin{figure}[htbp]
  \begin{center}
    \includegraphics[width=16cm]{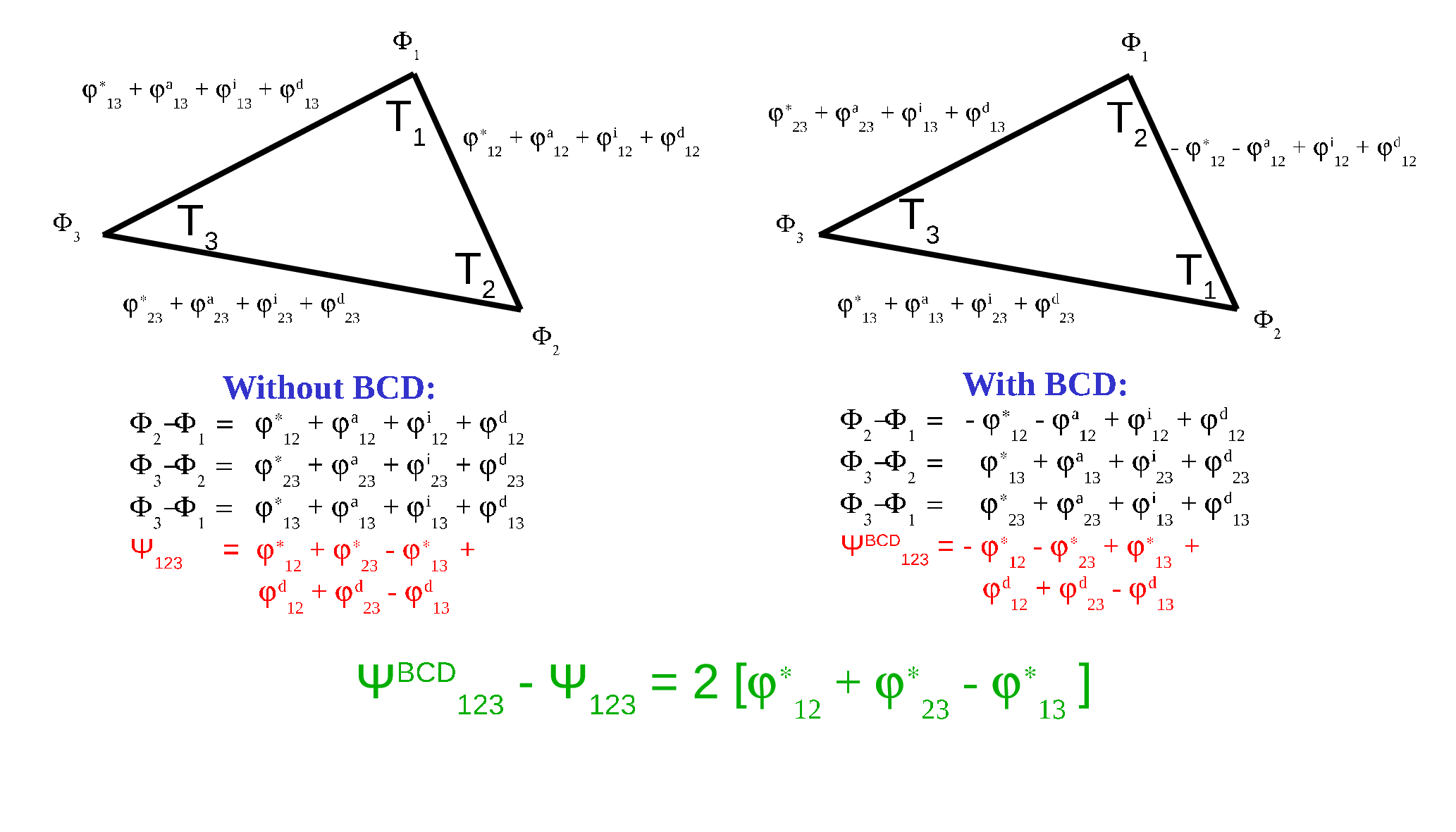}
  \end{center}
  \caption
  { \label{fig:closurePhasesCalib}
    The reason why closure phase can benefit from beam commutation is
    presented here: the instrumental effects taking place after the
    combination are cancelled out by subtracting the two closure
    phases recorded with and without BCD respectively.
  }
\end{figure}

Closure phase is insensitive to all atmospheric and most instrumental
effects, due to the fact that they cancel out in the closure
relation. However, the instrumental effects that are after the
recombination do affect the closure phase. These effects are in AMBER
mainly coming from camera distortion and from detector gain table
uncertainty. The Fig.~\ref{fig:closurePhasesCalib} illustrate how the
closure phase artifacts get calibrated out by the beam commutation.
Therefore, the same ideas apply when using the BCD with closure
phase.

\section{Experimental results}

We describe here measures of differential phase accuracy calibrated
both in standard way on a reference star and with the Beam Commuting
Device described above. We show and discuss some results in medium
resolution and low resolution in the K band (MR-K and LR-K). We also
show some very recent closure phase measures obtained on the ATs with
FINITO\cite{2008A&A...481..553L}, which allowed us for
the first time to obtain closure phases good enough for an accuracy
discussions. All our objects (but one) are calibrators supposed to
yield zero differential and closure phase, mostly obtained during the
commissioning with two ATs in July 2006 and during some Guaranteed
Time attempts to measure the very small phases and/or closure phases
indicating the presence of a planet around Tau Bootis. Please note
that in all figures, the values for the wavelength are only
indicative. An accurate spectral calibration was not relevant to the
problem discussed here and the wavelength scales can have offsets up
to 0.1$\mu$m. 

\subsection{Observations and data processing}

We illustrate the different observations and data processing steps on
the MR-K data obtained with two ATs in 2006. 
The data processing has been made with \texttt{amdlib} v2.0. The
exposures are processed individually, with the standard data
processinf and selection of 20\% of best frames. 
 \begin{itemize}
 \item In the STD mode, the chromatic OPD is fitted through and
  subtracted from the science and calibrator differential phases
  separately. Then, the calibrator differential phase is subtracted
  from the science one, yielding a calibrated differential phase. 
\item In the SPA mode, we compute the half difference between the
  differential phases obtained on BCD-out and BCD-in exposures. Then
  the chromatic OPD is fitted and subtracted yielding the calibrated
  differential phase.
\end{itemize}

\subsection{Differential phases calibration at medium spectral resolution}
\label{sect:spastd}

\begin{figure}[htbp]
  \begin{center}
    \includegraphics[width=14cm]{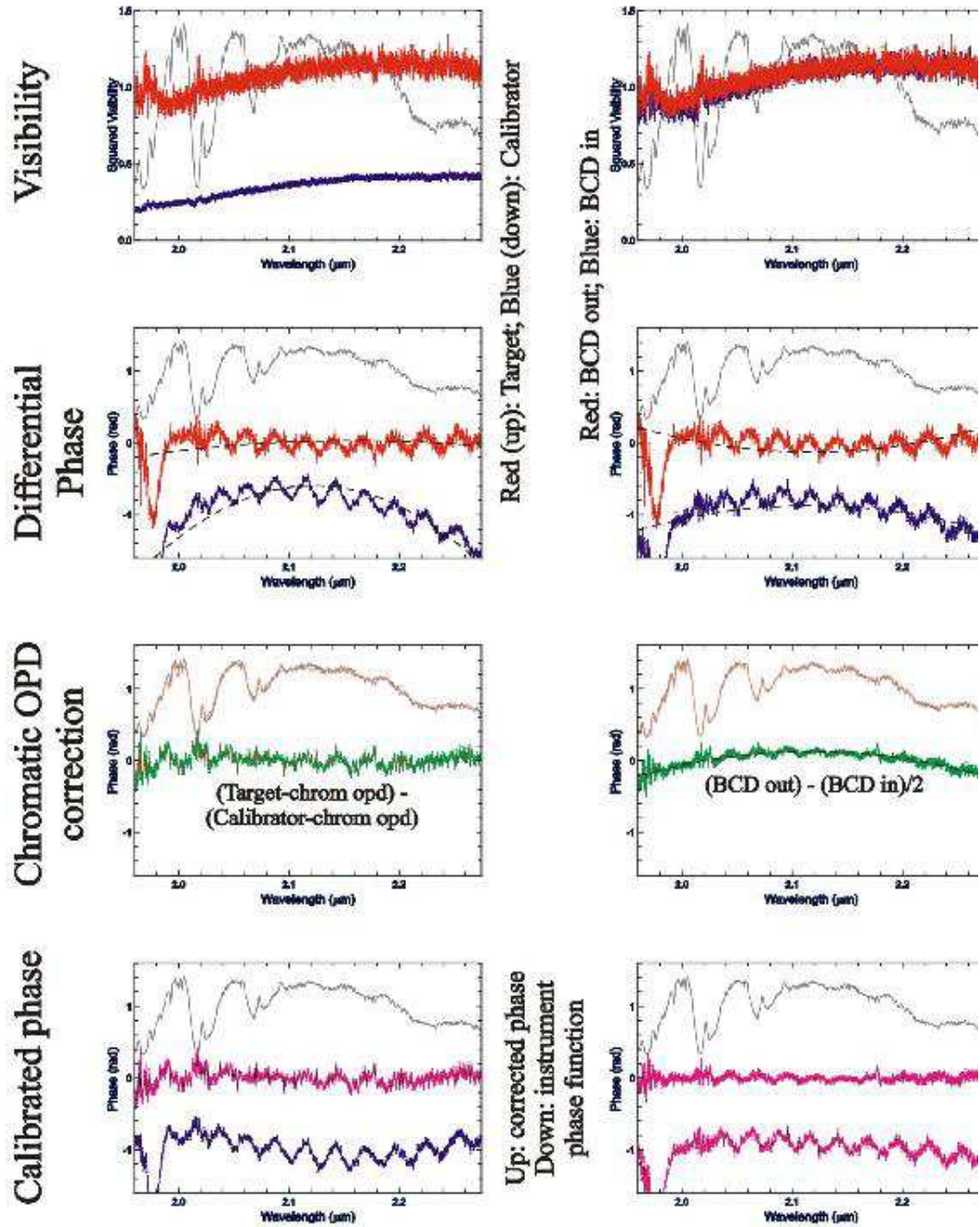}
  \end{center}
  \caption
  { \label{fig:phasesCalib}
    These plots show the comparison of standard calibration with a
    natural star (left column, STD) with the beam commutation
    calibration (right column, SPA stands for ``SPAtial
    modulation''). The top line shows the visibilities. One can see
    that the BCD does not affect the visibility measurement on the
    star (right column). Only the error bars are slightly larger to a
    flux loss of about 20-50\% (depending on how the BCD is
    aligned). The second line shows the raw differential phases for
    the star and the calibration star in the left part and for BCD-in
    and BCD-out for the right part. One can see the many effects
    affecting the differential phases, ranging from dichroic chromatic
    effects, ``socks'' and atmospheric longitudinal chromatic
    dispersion. Third line shows the ``calibrated'' differential phase
    (i.e. the difference of the two previous ones). One can see here
    the remaining artifacts: mainly atmospheric dispersion and
    ``socks'' residuals. The bottom line compares the final product,
    where the atmospheric dispersion was removed by fitting a 2nd
    order polynomial to the phase with the sum of the phases in the
    2nd line, giving an indication of the instrumental-only effects.
  }
\end{figure}

The Fig.~\ref{fig:phasesCalib} illustrate the different steps of the
standard calibration (left, labeled STD) and the calibration with the
BCD (right, labeled SPA, for SPAtial modulation, which as already
explained is actually what the BCD is approaching). The target is
Achernar (K=0.88, seeing~1.3 arcsec, t0=1.7 ms), 
and the calibrator is HD 219215 (K=0.1, seeing~1.5 arcsec). In spite
of the fast seeing, the data was acquired with single exposure time of
160 ms, allowing to have full K Band spectral window and more clear
results. The measured spectra is underlying in each figure. All figures
show the source + AMBER spectra in a fin grey line. The spectral
calibration has not 
been made with care since it does not affect the measurements, and
there is about 0.1$\mu$m offset in the wavelength labels. However, the
spectra of science and calibrators have been very carefully corrected
for any relative offset.

\begin{itemize}
\item Visibility: the two upper figures show the absolute visibility
  of the source (red) and of the calibrator (blue). Note that in the
  SDT mode the calibrator is quite resolved but much brighter than the
  source, explaining the much smaller error bars. In the SPA mode, the
  red curve stands for BCD-out and the blue for BCD-in. \textbf{We see
    that the BCD does not affect the absolute visibility value nor
    accuracy.} 
\item Differential phase: the second line of figures show the
  differential phases (red: science or BCD-out; blue: calibrator or
  BCD-in). The underlying dashed curve is a fit of the chromatic
  OPD. There are several stringent features. 
  \begin{itemize}
  \item The differential phase is affected by biases larger than the
    fundamental noise represented by the error. 
  \item At 1.95$\mu$m we see a strong phase distortion, larger than 1
    radian, and due to the K band dichroic. 
  \item There is a 0.3 radians modulation of the differential phase
    with a period of about 0.02$\mu$m. 
  \item The chromatic OPD is the same with the BCD-in and the BCD-out,
    but  for its sign, while it is substantially different for the
    science and the calibrator. 
  \end{itemize}
\item In the third line of figures we see the result of the BCD
  calibration at the right. Normally we should be left only with the
  chromatic OPD, which is indeed dominating the figure, and with
  stellar structures if any. Actually there is a clear signature of
  "rotating disk type" at the $Br_\gamma$ position (the source was the
  bright Be star Achernar). However, we see that the wavelength
  modulation has been very strongly reduced but not completely
  canceled. This means that the wavelength modulation is partially
  evolving faster than the 60 seconds BCD calibration cycle. 
\item Chromatic OPD: The BCD calibrated phase is clearly dominated by
  the chromatic OPD, which allows a clean fit of this feature. Then a
  final subtraction yields the fully calibrated differential phase in
  the lower right figure (green line). In the STD mode, the
  instrumental features corrupt the chromatic OPD fit and there is a
  clear chromatic OPD residual in the calibrated data. 
\item Calibrated differential phase: the lower line of figures show
  the differential phases calibrated by the standard method (left) and
  by the BCD method (right). The latest is clearly superior, although
  not perfect. 
  \begin{itemize}
  \item The residual of wavelength modulation are stronger and less periodic in the STD calibration. This indicates that the modulation variation is substantially larger other the much longer standard calibration cycle.
  \item The STD calibrated curve is contaminated by a chromatic OPD residual
  \item The STD calibrated curve shows in addition to the wavelength modulation many spikes and an apparently much larger noise. This cannot be noise, since we see that the raw calibrator differential phase is much less noisy, in terms of fundamental noise indicated by the error bars, than the BCD-in phase. This shows that the BCD eliminates artifacts that are currently dominated by the wavelength modulation but which will be dominant when the polarizer are replaced.
  \end{itemize}
\end{itemize}
In this work, Achernar was expected to behave as a bright calibrator (previous, standard, observations failed to detect a feature in the $Br_\gamma$ line) and we were quite surprised to see a clear disk signature in the SPA calibrated data. The same features actually exists in the STD differential phase but would be impossible to believe.

The table~\ref{table:obsLog} shows a more quantitative analysis of this data. We see that the SPA calibration gets much closer to the fundamental noise $\sigma$ due to photon, detector and background noise. More important the peak to valley features in wavelength are reduced by almost a factor 3 and are much more regular in SPA mode, allowing further fits in the continuum to be extrapolated in the line.

However, the BCD does not allow to cancel completely the wavelength modulation, which still prevents us reaching our highest goals in differential phase accuracy.

The available data in MR-K does not allow to conclude about the ultimate phase accuracy, since we were already too close to the fundamental noise of about 0.03 radians, but it seems clear that going much better in MR-K (with FINITO for example) for science applications such as Doppler imaging or asteroseismology would have to wait and hardware correction of the main source(s) of wavelength modulation.

\begin{table}[htbp]
  \centering
  \caption{
    \footnotesize{
      Table presenting the main comparison between the standard (STD)
      and beam-commuted (SPA) calibrations. The raw phases show
      typical patterns of  0.1-0.\,radians amplitude, due to the
      chromatic air dispersion as well as the ``socks'' and the
      instrumental chromatic effects. The standard calibration
      diminishes these $\lambda$-patterns basically to 3/2 of the
      fundamental noise ($\sigma$ column), whereas the
      beam-commutation allow to reach the fundamental noises.
    }
  }
  \label{table:obsLog}

  \begin{tabular}{|c|c|c|c|c|c|c|c|c|c|c|c|c|}
    \hline
    $\sigma$ & PTV ($\lambda$) & RMS ($\lambda$) & Comment \\
    \hline
    \multicolumn{4}{|c|}{Averaged phases}\\
    \hline
    0.033 & 0.34 & 0.172 & Achernar (BCD in)\\
    0.035 & 0.35 & 0.173 & Achernar (BCD out)\\
    0.025 &  0.34 & 0.294 & Calibrator (HD219215)\\
    \hline
    \multicolumn{4}{|c|}{Calibrated phases}\\
    \hline
    0.041 &  0.29 & 0.069 & STD \\
    0.048 &  0.12 & 0.044 & SPA\\
    \hline
  \end{tabular}
\end{table}

\subsection{Differential phase calibration in low spectral resolution}
\label{sect:phiLR}
\begin{figure}[htbp]
  \begin{center}
    \includegraphics[width=14cm]{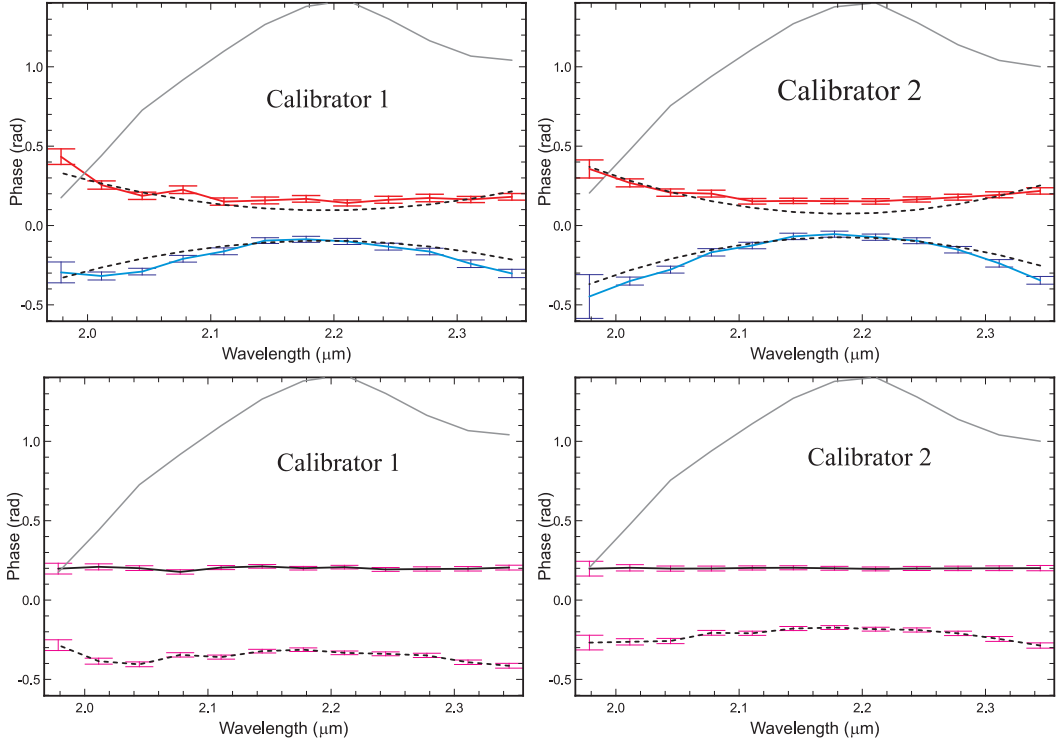}
  \end{center}
  \caption
  { \label{fig:phasesLR_AT_full}
    Differential phases using the BCD, for two calibrators observed one hour apart . In each plot, the upper thin black curve represents the spectrum. In the upper plots,
    the red and blue curves represent the differential phases observed, respectively, with and without the BCD. In their half-sum, the instrumental effects are in principle removed
    and only the chromatic OPD effects should remain. These have been approximated using a polynomial fit, which is represented by the thin black curves superimposed on the differential
    phases curves in the upper plot. The lower plots show the half-differences (solid line) and half-sum (dashed line) of these curves, after subtraction of the chromatic OPD fit.
    Note that the BCD-calibrated phase, which appears in the half-difference (BCD out - BCD in)/2, is indeed very much flatter and smoother than the phase observed without BCD.
  }
\end{figure}

Another set of data shows the effect of the BCD calibration with observations made in low spectral resolution ($R=35$).
The results appearing in Fig. \ref{fig:phasesLR_AT_full} illustrate the effect of the BCD calibration: after subtraction of the
chromatic dispersion fit, the calibrated phase is flattened and smoothed from instrumental artifacts.

Table \ref{table:results_LR_ATs} allows to compare the BCD calibration with the standard calibration, consisting
in making the difference Target - Calibrator.

The key result is that the STD calibration leave PTV wavelength features larger than 0.15 radians, corresponding to the variation of the LR instrumental differential phase in one hour, while the SPA calibration reduces the PTV features in lambda below the statistical RMS between frames in a given spectral channel, which we used to consider as a good estimate of the fundamental noise. With this limited set of poor quality data it is possible to conclude that the SPA mode allows to reach a differential phase accuracy better than 0.01 radians.

This measure was also intended to check if calibration combining SPA calibrated differential phases from a source and a calibrator could allow further progress, for example by removing systematic effects introduced by the BCD, but this tests results impossible since both calibrators are cleaned below the fundamental noise limit.

\begin{table}[htbp]
  \centering
  \caption{
    \footnotesize{
      Summary of the statistics on the differential phase for a science source and
      its calibrator, in different observation and calibration modes: for BCD OUT and IN, and
      for a calibration using the BCD (phase OUT - phase IN)/2 or using the standard Target-Calibrator
      comparison. In each case, we indicate the statistical RMS (averaged over $\lambda$) between the exposures,
      and the Peak-to-Valley (PTV)  and RMS over $\lambda$ of the averaged phase. For the latter two quantities,
      the polynomial fit of the chromatic dispersion, estimated for the average BCD-calibrated phase,
      was previously subtracted in the same way as presented in Fig. \ref{fig:phasesCalib}.
    }
  }
  \label{table:results_LR_ATs}
  \begin{tabular}{|c|c|c|c|}
    \hline
    Source and  & Statistical RMS  & PTV  over $\lambda$ & RMS over $\lambda$ \\
    calibration mode & between exposures & & \\
    \hline
    Calibrator 1 (OUT) & 0.011 & 0.15 & 0.0288\\
    Calibrator 1 (IN) & 0.018  & 0.15 & 0.027\\
    Calibrator 2 (OUT) & 0.015 & 0.10 & 0.027\\
    Calibrator 2 (IN) & 0.019 & 0.10 & 0.026\\
    Cal 1 - Cal 2 (STD) & 0.012 & 0.15 & 0.011 \\
    Calibrator 1 (SPA) & 0.017 & 0.01 & 0.002 \\
    Calibrator 2 (SPA) & 0.013 & 0.03 & 0.009 \\
    \hline
  \end{tabular}
\end{table}

\subsection{Closure phase calibration in low spectral resolution}

\label{sect:closureLR}

\begin{figure}[htbp]
  \begin{center}
    \includegraphics[width=14cm]{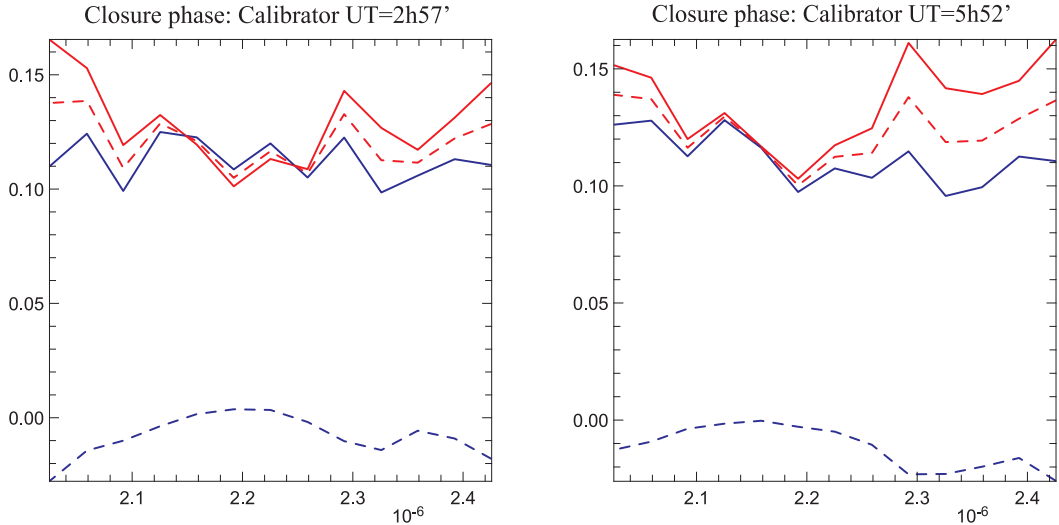}
  \end{center}
  \caption
  {
    \label{fig:CP_BCD}
    Closure phases observed in Low Resolution with the ATs, using the Beam Commuting Device (BCD), for two calibrators. Full lines refer to averaged exposures
    without (blue) and with (red) the BCD.
    The red dashed line is the half-sum of the closure phases (with and without BCD) and represent, in principle, the instrumental contribution to the non-calibrated closure phase.
    The blue dashed line  is the half-difference of the closure phases, where the instrumental contribution should have been corrected by the use of the BCD. We note that
    these calibrated closure phases are indeed improved, with a lower average level and a smoother shape. But that they still show a significant residual, the
    theoretically expected closure phase for these targets being a flat null curve.
  }
\end{figure}

The principle of closure phase calibration using the Beam Commutation
Device, described in Sec. \ref{sbsect:CPCalib},  
is illustrated in Fig. \ref{fig:CP_BCD}. This is the result of some
recent observations made in low spectral resolution with FINITO, which
gave us for the first time high quality closure phases. We had, for 
calibrators presented here, a series of 10 or 15 exposure pairs, each
of them containing an observation with and without the BCD. The
half-difference of the closure phase within each pair
should in principle be corrected from all the OPD effects affecting
each beam, and from the post BCD instrumental effects if these are
stable over the BCD cycle period. We find that the BCD corrects the
general closure phase bias, smooths channel to channel closure phase
features but remains affected by small time variable effects which can
be the LR aspect of the wavelength modulation on the closure phase.
\begin{itemize}
\item The raw closure phases, both with and without the BCD, are
  globally offset by about 0.15 radians. This is probably due to the
  obsolescence of the instrument calibration, made about three hours
  before the first observation, even if the offset change by much less
  than 0.1 radians during the three hours separating the two displayed
  observations. The BCD calibration brings the average closure phase
  between 0.01 and 0.02.
\item The closure phases show features in each spectral channels,
  which are partially stable in pattern but quite variable in
  amplitude (by about 0.02 radians in 3 hours).
\item The BCD calibrated closure phase is much smoother in wavelength,
  with a PTV of about 0.02 radians and a RMS of about 0.008
  radians. This RMS remains substantially larger than the fundamental
  noise in that case (0.002 radians). So we are limited by the
  instrument at about 0.01 radians per calibrated observations, at
  least with the current hardware and software.
\end{itemize}

As a whole, the BCD-calibrated closure phases look much more to what we would expect from the astrophysical signature of this observable (i.e., in the present case
with calibrators, a flat curve at 0). Nevertheless, both the average and the lambda-rms levels of our calibrated closure phases are well above effects due to the fundamental
noise. This therefore indicates that the calibration is not complete. The hypothetical causes for this residual may well be some parasite Fabry-Perot interferences
within AMBER dioptres and/or some non-mono-mode effect of the spatial filtering, resulting in the ``socks'' pattern already mentioned in Sec. \ref{sect:spastd}.

\section{Conclusion}
We have shown that the AMBER differential and closure phases display
evolving patterns much larger than expected and specified. With a
standard calibration using a reference star, it seems difficult to
trust phase features smaller than 0.1 radians. This is confirmed by
data on other stats and in other bands (MR-H and LR-H) that we could
fit in this paper. Since our observations, we have learned that the
main perturbation results from Perrot Fabry effects in various glass
and air blades, both within AMBER and before it (IRIS feeding
dichroic). The major cause for this perturbations has been proved to
be the AMBER polarizing filters which will be soon replaced and
relocated to minimize this wavelength beating effects. 

The specific phase calibration function that we have developed improves the situation but quite not enough. With our Beam Commutation Device we can approach the 0.01 radian limit in the best cases, but this remains much larger than the fundamental noise limits, which are at least 10 times smaller with the ATs and 100 times smaller with the UTs.

The BCD clearly improves other sources of phase errors, more likely to be found in the calibration of the detector or of the position of the beams on the detector. This gives us good hope that the combination of Perrot Fabry effect suppression and of BCD calibration will allow us to approach the 0.001 radians limit, that we already approach in LR-K differential phase.

With the current hardware and without using the BCD, we had 0.1 radian (5 degrees) accuracy on phase, which was enough for the first science application on large disks ($\alpha$ Arae cite{}) or on binaries with almost equal components ($\gamma^{2}$ Velorum cite{}). With the BCD and a carefully data processing, we can approach the 0.01 radian specification sufficient for most circumstellar physics (disks and wind in stars of all ages) and by combining hardware and software upgrades with the BCD calibration we still confidently hope that the milliradian limit needed for extra solar planets and stellar activity work will be achieved.


\bibliography{biblio}   
\bibliographystyle{spiebib}   

\end{document}